\documentclass[aps,prd,12pt,preprintnumbers,nofootinbib,superscriptaddress]{revtex4}
\usepackage{graphicx}
\usepackage{mathrsfs}
\usepackage{amssymb,amsmath}
\def\beq{\begin{equation}}
\def\eeq{\end{equation}}

\newcommand{\bo}{\raise-1mm\hbox{\Large$\Box$}}

\newcommand{\f}[2]{\frac{#1}{#2}}

\newcommand{\la}{\langle}
\newcommand{\ra}{\rangle}
\newcommand{\w}{\omega}
\newcommand{\kp}{\kappa}
\newcommand{\be}{\begin{equation}}
\newcommand{\ee}{\end{equation}}
\newcommand{\bea}{\begin{eqnarray}}
\newcommand{\eea}{\end{eqnarray}}

\begin{document}

\title{Black Hole - Moving Mirror II: Particle Creation}
\author{Michael R.R. Good}

\address{Physics Department, Nazarbayev University,\\
Astana, Republic of Kazakhstan\\
E-mail: michael.good@nu.edu.kz}

\author{Paul R. Anderson}

\address{Department of Physics, Wake Forest University,\\
Winston-Salem, North Carolina  27109, USA\\
E-mail: anderson@wfu.edu}

\author{Charles R. Evans }
\address{Department of Physics and Astronomy, University of North Carolina,\\
Chapel Hill, North Carolina 27599, USA\\
E-mail: evans@physics.unc.edu }

\begin{abstract}
There is an exact correspondence between the simplest solution to Einstein's equations describing the formation of a black hole and a particular moving mirror trajectory.  In both cases the Bogolubov coefficients in 1+1 dimensions are identical and can be computed analytically.  Particle creation is investigated by using wave packets. The entire particle creation history is computed, incorporating the early-time non-thermal emission due to the formation of the black hole (or the early-time acceleration of the moving mirror) and the evolution to a Planckian spectrum.
\end{abstract}


\maketitle


\vspace{0.5cm}
In the previous paper\cite{paper1}, hereafter called Paper 1, it was shown that there is an exact correspondence between the particle production that occurs for a massless minimally coupled scalar field in a (1+1)D spacetime in which a black hole forms from the collapse of a null shell
and the particle production for this field which results from an accelerating mirror with a particular trajectory in (1+1)D.
Here we investigate the time dependence of the particle production process for these cases and the approach to a thermal distribution at late times.
We also compute the time dependence of the energy flux for the field in the mirror spacetime.

In the black hole case
the trajectory is along the ray $v=v_0$ with $v = t + r$ in the flat space region inside the shell and $v = t_s + r_*$ in the Schwarzschild geometry outside the shell with $t_s$ the
time in the usual Schwarzschild coordinates and
$r_* \equiv r + 2 M \ln \frac{r-2M}{2M} $.  A Penrose diagram for the spacetime is given in Fig. 1 of Paper 1.  The mirror trajectory is given by
\be z(t) =  v_H -t - \f{W(2 e^{2 \kappa(v_H - t)})}{2\kappa} \;,  \label{trajectory} \ee
 with $W$ the Lambert W function.
It is plotted in Fig. 1 for the case $v_H = 0$.  It begins at $z = \infty$ at rest and asymptotically approaches the ray $v = v_H$.
At late times this trajectory is similar to that studied by Carlitz and Willey\cite{Carlitz:1986nh} for which the particles are at all times created in a thermal distribution.\footnote{See Good, Anderson, and Evans\cite{Good:2013lca} for the explicit $z(t)$ trajectory.}  However, at early times the mirror we consider has a very different trajectory and, as shown below, the particles are only created in a thermal distribution at late times.

\begin{figure}[ht]
\begin{minipage}[b]{0.45\linewidth}
\centering
 \rotatebox{90}{\includegraphics[width=\textwidth]{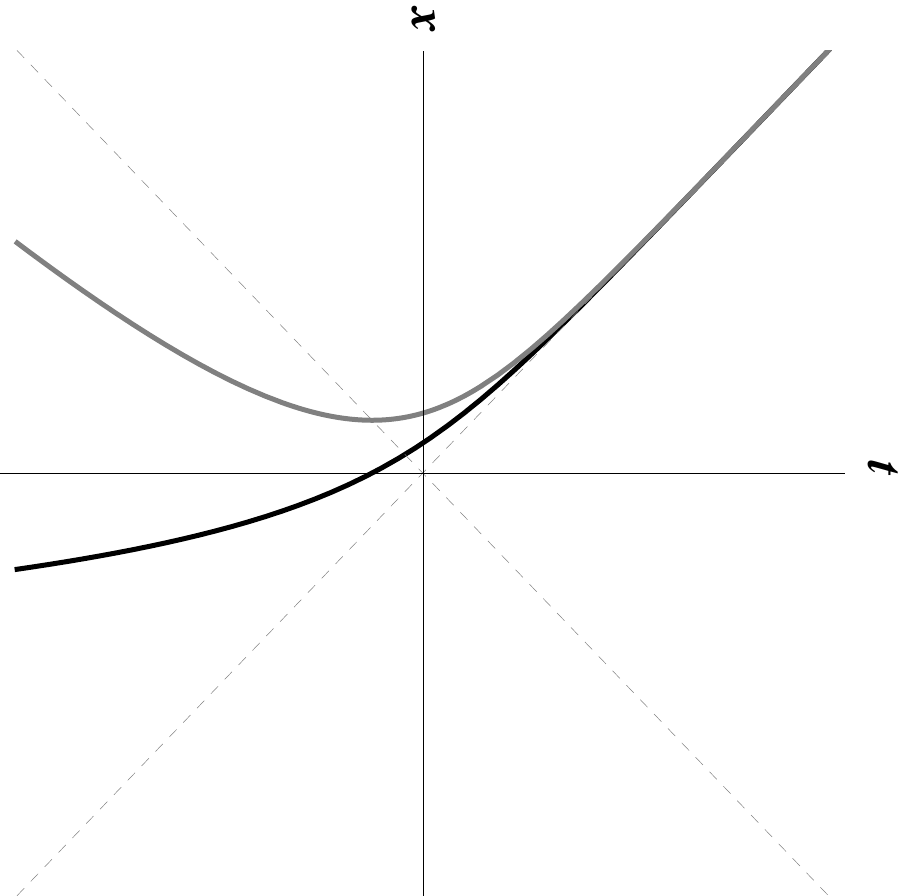}}
\caption{\label{fig:omexST} The new trajectory~\eqref{trajectory} with $v_H = 0$ and $\kp=2$ is shown (right) along with the Carlitz-Willey trajectory with the same asymptotic behavior (left).}
\end{minipage}
\hspace{0.5cm}
\begin{minipage}[b]{0.45\linewidth}
\centering
\includegraphics[width=\textwidth]{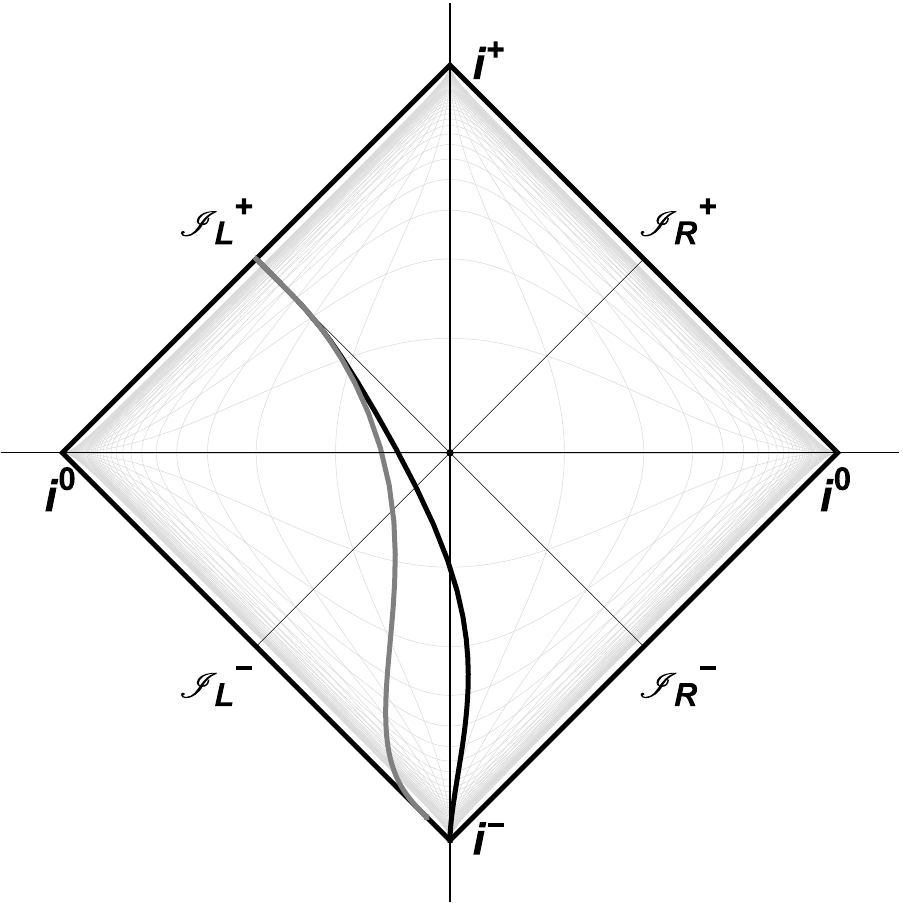}
\caption{\label{fig:omexPR} Penrose diagram showing the new mirror trajectory~\eqref{trajectory} with $v_H = 0$ and $\kp=2$ (right) and the Carlitz-Willey trajectory (left).  }
\end{minipage}
\end{figure}


  In Paper 1 the Bogolubov coefficients between the ``in'' and ``out'' states were computed for the modes in the ``out'' state which go to $\mathscr{I}^{+}_R$.
  The one relevant to particle production can be written in the form
      \be      \beta_{\w \w'} =  -\frac{ \sqrt{\w \w'}}{2 \pi \kappa (\w +\w')} \, e^{-i(\w+\w') v_H} \,e^{- \pi \w/(2 \kappa)} e^{-i \w/\kappa \ln [ (\w + \w')/\kappa]} \, \Gamma(i\w/\kappa) \;.  \label{beta} \ee
The total number of particles produced with frequency $\w$ is
\be  N_\omega = \int_0^\infty d \omega' |\beta_{\w \w'}|^2  \;,  \ee
and
\be |\beta_{\w\w'}|^2 = \f{\w'}{2\pi\kp(\w+\w')^2} \f{1}{e^{2\pi \w/\kp} - 1} \;. \ee
Thus an infinite total number of particles is produced.  However, if the time
dependence of the particle production is investigated then a finite number
of particles is produced in any finite time interval\cite{Hawking:1974sw}.

To investigate the time dependence of the particle creation process we construct localized wave packets which were used by Hawking\cite{Hawking:1974sw} and carried out for specific accelerating mirror models by Good-Anderson-Evans\cite{Good:2013lca} and Good-Ong\cite{Good:2015nja}.  When applied to the mode functions, the packets form a complete orthonormal set.
The packets can also be applied directly to the Bogolubov coefficients\cite{Fabbri:2005mw} with the result that
\be
\beta_{jn\w'} =
\f{1}{\sqrt{\epsilon}}\int_{j\epsilon}^{(j+1)\epsilon}
d\w \; e^{\frac{2\pi i \w n}{\epsilon}} \beta_{\w\w'} \;.
\label{beta-packet}
\ee
The value of $n$ is related to the time and the value of $j$ is related to the frequency of the packet.  The average total number of particles produced  in a given packet which arrive at
$\mathscr{I}^{+}_R$ is
\bea
\la N_{jn}\ra &=& \int_0^\infty d\w' |\beta_{jn,\w'}|^2   \;.
\label{Njn}
\eea
The particles arrive at $\mathscr{I}^+_R$ in the range of frequencies $j \epsilon \leqslant \omega \leqslant (j+1) \epsilon$ and in the range of times $ (2\pi n - \pi)/\epsilon \lesssim u \lesssim (2 \pi n + \pi)/\epsilon$.  

\begin{figure}[ht]
\begin{minipage}[b]{0.45\linewidth}
\centering
\includegraphics[width=\textwidth]{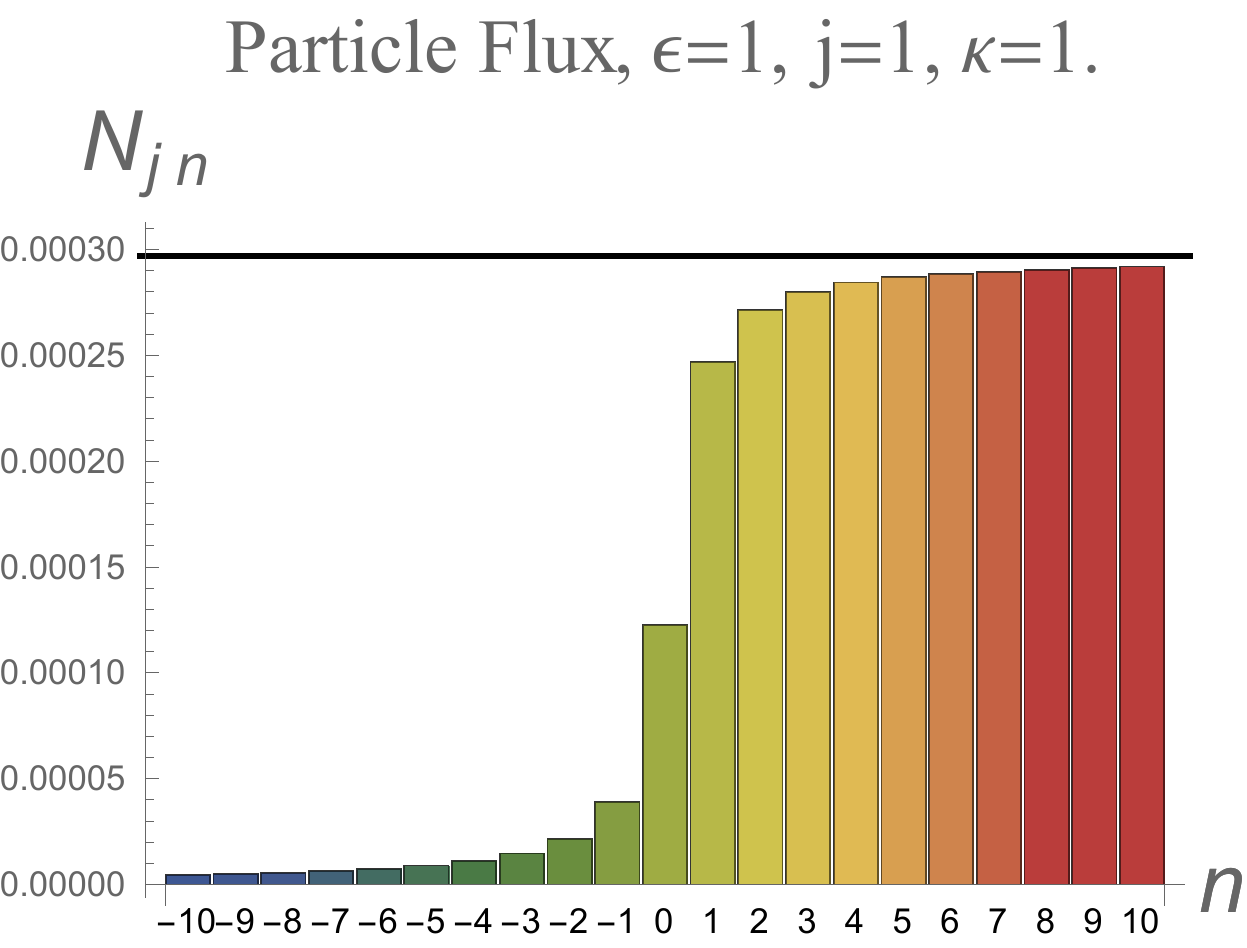}
\caption{\label{fig:particleflux} Number of particles as a function of $n$ in the frequency bin with $j =1$ and $\epsilon = 1$.  The black line is
the corresponding number of particles produced for the Carlitz-Willey trajectory.}
\end{minipage}
\hspace{0.5cm}
\begin{minipage}[b]{0.45\linewidth}
\centering
\includegraphics[width=\textwidth]{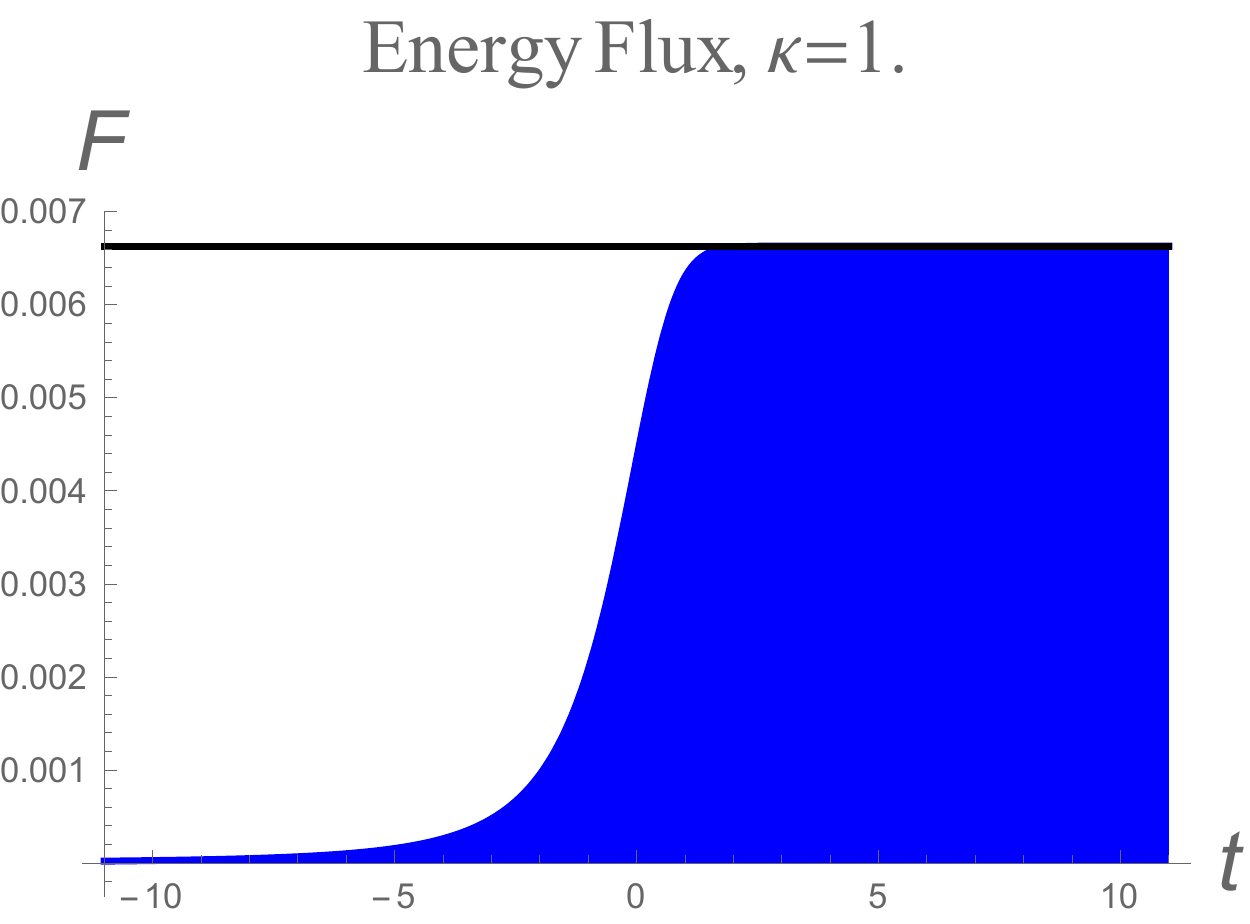}
\caption{\label{fig:energyflux}Energy flux of a quantized massless minimally coupled scalar field for the accelerating mirror spacetime for $v_H = 0$.  The black line is
the energy flux for the Carlitz-Willey trajectory.}
\end{minipage}
\end{figure}

A plot of the time dependence of the particle production for the frequency band $\epsilon = j = 1$ is given in
 Figure (\ref{fig:particleflux}).  We have found that it is not possible to simultaneously obtain both good time resolution
 and a good frequency spectrum.  For good time resolution the bins must be wide enough in frequency that one has a coarse-grained frequency spectrum.
 For a good frequency spectrum the bins must be long enough in time that one has poor time resolution.  This was also found
 for all of the trajectories investigated  in Ref.~\cite{Good:2013lca} except the Carlitz-Willey trajectory which gives a time independent spectrum.

If Eq.~\eqref{beta} is substituted into Eq.~\eqref{beta-packet} then in the late time, large $n$ limit one can see that the dominant contribution
to the integral comes from values of $\w'$ for which the arguments of the oscillating exponentials cancel or nearly cancel and which
therefore satisfy the condition $\w' \gg \w$.
In this limit
\be |\beta_{\w\w'}|^2 \sim \f{1}{2 \pi\kp\w'} \f{1}{e^{2\pi \w/\kp} - 1} \;,
\ee
and one sees that there is a thermal distribution of particles with temperature $T=\frac{\kappa}{2\pi}$.  Such a late time thermal distribution was found
 for black hole radiation in\cite{Hawking:1974sw} and for mirrors with a particular class of asymptotically null trajectories in\cite{Davies:1977yv} .

While particle production is the same for the collapsing null shell and accelerating mirror spacetimes, the stress-energy tensor for the scalar field will not be.
Here we compute it for the accelerating mirror spacetime.  
The general form of the energy flux for any mirror trajectory is\cite{Davies:1976hi}
\be\label{stressz} F(t) \equiv \la T_{uu} \ra = \f{\dddot{z}(\dot{z}^2-1)-3\dot{z}\ddot{z}^2}{12\pi(\dot{z}-1)^4(\dot{z}+1)^2},\ee
where the dots are time derivatives. The energy flux, using the trajectory of Eq.(\ref{trajectory}), as function of time gives a result in terms of the Lambert W function:
\be F(t) = \frac{\kappa ^2 \left(2 W\left(2 e^{2 \kappa(v_H-t}\right)+1\right)}{3 \pi  \left(W\left(2 e^{2 \kappa (v_H- t}\right)+2\right)^4}. \ee
This is plotted as a function of time in Fig.~\ref{fig:energyflux}.  In the late time limit it approaches the thermal value given by the Carlitz-Willey trajectory at all times.


We have investigated the time dependence of the particle production that occurs in (1+1)D for a massless minimally coupled scalar field when a black hole forms from gravitational collapse of a null shell and when a mirror in flat space has the trajectory~\eqref{trajectory}.
Both the number of particles and the energy flux (in the mirror case) increase in time and approach the values they have for a thermal distribution.

\section*{Acknowledgments}

MRRG thanks Xiong Chi at the Institute of Advanced Studies in Singapore, where a one-to-one correspondence was initially discussed in terms of an exact expanding cosmological spacetime-moving mirror (CS-MR).  This work was supported in part by the National
Science Foundation under Grant Nos. PHY-0856050, PHY-1308325, and PHY-1505875 to Wake Forest University, and PHY-1506182 to the University of North Carolina, Chapel Hill.

\end{document}